\documentstyle[12pt]{article}
\author{Hans - J\"urgen Schmidt} 
\title{A two-dimensional representation of
           four-dimensional gravitational waves}
\date{}
\begin{document}
\maketitle

\newcommand{\be }{\begin{equation}}
\newcommand{\ee }{\end{equation}}

\centerline{
Universit\"at  Potsdam, Inst. f. 
Mathematik}
\centerline{
 D-14415 POTSDAM, PF 601553, Am Neuen Palais 10, Germany}

\begin{abstract}
The 
Einstein equation in $D$ dimensions, if restricted to the
class of 
 space-times possessing  $n  =  D - 2$
commuting hypersurface--orthogonal  
Killing vectors, can be 
  equivalently written as  
metric-dilaton gravity in 2 dimensions
  with $n$ scalar fields.

\bigskip

For $n=2$, 
this  results reduces to the known reduction 
of certain
 $4$-dimensional metrics which include gravitational waves. 
 Here, we give such a representation which leads to a new
proof of the Birkhoff theorem for plane-symmetric space--times, 
and which leads to an explanation, in which sense two 
(spin zero-) scalar fields in 2 dimensions may incorporate the 
(spin two-) gravitational waves in 4 dimensions. 
(This result should not be mixed up with well--known 
 analogous 
statements where, however, the 4--dimensional space--time  
is supposed to be spherically symmetric, and then, of course,
the equivalent 2--dimensional picture cannot mimic any
gravitational waves.)

\bigskip 

Finally, remarks on hidden symmetries in 2 dimensions are made.


\end{abstract}

\medskip


     04.60Kz lower-dimensional models

\bigskip

 {\it Int. J. Mod. Phys. D in print}

\newpage 

\section{Introduction}

Gravitational waves and their collisions in a Friedmann 
 universe have been discussed recently $[1]$ by use of the metric
\begin{equation}
ds^{2} = e^{M}(dt^{2} - dz^{2}) - e^{\phi}
 (e^{\psi}dx^2 + e^{-\psi} dy^{2}) \ee 
where $M, {\phi}$ and ${\psi}$ depend on $t$ and $z$ only.
The space-times possessing a metric of type eq. (1)
can be invariantly defined by requiring 
 that two commuting hypersurface-orthogonal 
space-like Killing vectors exist.

In  higher-dimensional cosmology,
 $D$-dimensional  space-times possessing  $n \, = \, D - 2$
commuting hypersurface--orthogonal space--like  
Killing vectors are discussed, 
 cf. [2, 3] and refs. cited there. They can 
be expressed via 
\be
ds^{2} = d {\sigma}^{2} -\sum _{k=1}^{n}  a_{k}^{2}
 (x^{\alpha})(dx^{k})^{2} \ee
where 
\be d {\sigma}^{2} = g_{\alpha \beta} (x^{\alpha}) 
dx^{\alpha}dx^{\beta} \ee
$\alpha, \beta \in \{ 0,n+1 \}, 
\  a_k>0,$ 
 and $g_{\alpha \beta}$ has signature $(+ -)$.

For $D=4$, metric (2, 3) can locally be written in the 
form of eq. (1) \ ($x^{0}=t, \ x^{3} = z$), because 
the 2-dimensional space--time $d\sigma^2$ is conformally
 flat, at least locally.

In the last two years, much progress, cf. refs. $[4-15]$,
 has been made
in dealing with $2$-dimensional gravity, both
classically and with its 
quantization. However, this progress lacks from a satisfying
physical application due to the hypothetic nature of 
(super)-strings. So, it is often considered as toy model to study
conceptual features of gravity under simplified circumstances. 
For instance, the evaporation of $1+1$-dimensional black holes
[13] should tell something about the evaporation 
of $3+1$-dimensional ones. Analogously, the collapse of massless
scalar fields in $1+1$-dimensional dilaton gravity [14] 
should be similar in structure to black hole formation in 
the $3+1$-dimensional case.

In the present paper, however, we go beyond a toy model:
known results from 2-dimensional gravity shall be applied to
 higher--dimensional models under such circumstances where the
correspondence between the low and the high dimension can be
given explicitly.

In section $2$, the $D$-dimensional Einstein equation
 for the metric (2, 3) is rewritten in a 
 2-dimensional form, in  sct. $3$ 
we describe the peculiarities for 
 $D = 4$,  
and sct. 4 discusses the results. We enclose an appendix
 on the hidden symmetries in 2 dimensions. 

\bigskip

\section{The $D$-dimensional Einstein equation}

 The Ricci tensor of $ds^2$, eq. (2) will be  denoted by 
 $R_{AB}$,  \ $A,B$ take the values $0,..., \, n+1$. 
 The Einstein sum convention shall be applied to
indices 
$\alpha, \beta$ and $A,B$ but not to $i,j,k$.
 Let $\phi _{k} = \ln a_{k}, \phi = \sum^{n}_{k=1}
\phi_{k}$, 
and $\psi_{k}=\phi_{k} - \frac{\phi}{n}$.
 Consequently 
\be \sum _{k=1}^{n} \psi_k =0 \ee
and eq. $(2)$ can be rewritten as
\be
ds^{2} = d {\sigma}^{2} - e^{2\phi / n} \sum_{k=1}^{n} 
e^{2 \psi_k} (dx^{k})^{2} \ee
where $\phi$ and $\psi _k$ depend on the $x^{\alpha}$ only.

The
Ricci tensor of $d {\sigma}^{2}$ is denoted by 
$P_{\alpha \beta}$, and then the non-vanishing 
components of
 $R_{AB}$  are given by
\be R_{\alpha \beta} = P_{\alpha \beta} - \phi_{; \alpha \beta}
- \frac{1}{n} \phi_ {, \alpha} \phi_{, \beta}-
\sum _{k=1}^{n}
\psi_{k, \alpha} \psi_{k, \beta} \ee
and 
\be R_{kk}= ( \Box \phi_{k} + \phi_{k}^{, \alpha} 
\phi_{, \alpha}) e^{2 \phi_k} \ee
With $G = \vert \det g_{AB} \vert$ and 
$g= \vert \det g_{\alpha \beta} \vert $ we get 
$G = ge^{2 \phi}$, so
that eqs. $(6,7)$ together give the Einstein-Hilbert
 Lagrangian in $D$ dimensions 
 for metric $(5)$ as
\be L=R \sqrt{G}=(P- \frac{n-1}{n} \Box \phi -
\sum_{k=1}^{n} \psi_{k, \alpha} \psi_{k}^{, \alpha})e^{\phi}
\sqrt{g} \ee
It should be noted that we added a suitable multiple of the
divergence 
$(e^{\phi} \phi^{, \alpha})_{, \alpha}$
 to get this simple equation.

It holds (this is a non-trivial statement): The variational
derivatives of $L$ eq. (8) with respect to
 the 2-dimensional 
 metric 
$g_{\alpha \beta}$ and $\phi$ and those $\psi_k$ fulfilling the
constraint eq. (4)
 lead to the  $D$-dimensional Einstein equation
for metric $(5)$. That every $D$-dimensional Ricci--flat space
fulfils these conditions is trivial. The reverse statement,
however, is non-trivial. Its
 proof  uses the fact that by this procedure, 
 no spurious 
solutions can appear.\footnote{Spurious solutions will 
appear e.g. if
we restrict to synchronized time $t$ before the
variation with respect to $g_{\alpha \beta}$ is carried out.} 
 
\noindent 
Of course, the variation of $R \sqrt{G}$ with respect to 
$g_{AB}$ gives the  $D$-dimensional Einstein tensor.

The next problem is how to deal with the constraint 
eq. $(4)$. One
could simply mention that it is harmless, because its validity
 in the initial conditions implies its validity 
 everywhere, but one can also do it explicitly by eliminating
of one of the fields $\psi_k$. For $n=2$ we introduce $f_1$ via 
$$\psi_1 = f_1/\sqrt 2 \, , \  \psi_2 = - f_1/\sqrt 2 $$
For $n=3$ we introduce $2$ scalars $f_{i}, i=1,2$ 
 (the Misner parametrization)  via 
 $$\psi_{1}=\frac{f_1}{\sqrt{6}} + \frac{f_2}{\sqrt{2}},
\qquad 
\psi_{2}=\frac{f_1}{\sqrt{6}}-\frac{f_2}{\sqrt{2}},
 \qquad 
\psi_{3}=- \frac{2f_1}{\sqrt{6}}$$
and get
$$\sum_{k=1}^{3} \psi_{k,\alpha} \psi_{k}^{, \alpha}=
\sum_{i=1}^{2} f_{i,\alpha} f_{i}^{, \alpha}$$
and for larger values $n$, analogous linear relations hold such
that eq. (4) is identically fulfiled, and, moreover,
\be
 \sum _{k=1}^{n} \psi _{k, \alpha} \psi_{k}^{, \alpha} =
\sum_{i=1}^{n-1}
f_{i,\alpha} f_i^{,\alpha}
\ee 
The comparison with eq. (1) and (5) for $n=2$ requires
$\psi=2\psi_1$.

Result: 
 Lagrangian $(8)$ with the $n+1$ scalar
fields $\phi, \psi_{k}$ subject to the constraint
 eq. $(4)$ is equivalent to the
Lagrangian
\be
L = (P-\frac{n-1}{n} \Box \phi - \sum_{i=1}^{n-1}
f_{i,\alpha} f_{i}^{,\alpha} ) e^{\phi} \sqrt{g} \ee
 with the $n$ scalar fields $\phi, f_{i}$.

This kind of reduction seems to be new, and it shall be applied
as follows: For 2-dimensional nonlinear gravity 
$L=f(P) \sqrt g$ a generalized Birkhoff theorem [16] 
states that every solution of the field equation 
possesses an isometry. In [17], nonlinear gravity is shown to be 
conformally equivalent (after suitable field redefinitions)
to dilaton metric theories in 2 dimensions following from 
\be
{\cal L}=   
 D(\phi) P - V(\phi) - Z(\phi)(\nabla \phi)^2 + 
Y (\phi) \Box \phi
\ee
so that 
this Birkhoff theorem holds also in all these 
theories, cf. [4, 9, 13]. In [18] it is explained, that this 
theorem is a consequence of the fact that the traceless part of
the
Ricci tensor identically vanishes in $2$ dimensions. Let us now
apply this Birkhoff theorem: If we restrict to 
solutions with vanishing fields $f_i$, then Lagrangian $(10)$ has
the structure $(11)$, and one isometry can be found. The 
vanishing of $f_i$
means that all functions $a_k$ in eq. $(2)$ coincide, and so we
may conclude:

If all functions $a_k$ in metric $(2)$ are equal, and 
$ds^2$ (2) is $D$-dimensionally Ricci-flat, then besides the $n$
isometries 
$$\frac{\partial}{\partial x^k},$$ 
one further 
isometry exists, its Killing vector\footnote{In case 
$\epsilon^{\alpha \beta} \, P_{, \beta} =0$ in a
whole region, then $P=$ const., and we have not only
one but three additional isometries.}
 is proportional to 
$\epsilon^{\alpha \beta} P_{,\beta}$, where 
$\epsilon^{\alpha \beta}$ is the Levi-Civita tensor in
$d \sigma^2$.
 
\bigskip

\section{The 4-dimensional Einstein equations}

If we restrict the results of sct. 2 to $D=4$, then we recover a
couple of known results. Nevertheless, 
 it proves useful to point them out in
our general approach.

The vanishing of the $f_i$ is equivalent to put  
 $\psi = 0$ in eq. (1), i.e., one has  
plane-symmetric space--times. The Birkhoff theorem
for them has already been proven in [19, 20]. 
Our approach has the advantage that the 
additional Killing vector can be explicitly given 
without the need to specify a coordinate system.
The analogous procedure for spherical instead
of plane symmetry can be found in [21].

Gravitational waves in our Universe -- even if not yet 
surely identified -- represent a serious object of
experimental undertakings. Therefore, metrics like eq. (1)
deserve to be carefully understood also theoretically. 
The controversy [22, 23] about ``what can be 
learned from metric--dilaton gravity in two
 dimensions for the theoretical understanding 
of gravitational waves in four dimensions ?''
 can now be resolved by mentioning the above deduced 
equivalence of the 4-dimensional Einstein equation 
for eq. (1) to the 2-dimensional metric dilaton theory eq. (10)
for $n=2$. 

Let us add some related results: In chapter 15 of [24] the
reduction of metric (1) to a 2-dimensional formulation is given
(and, by the way, without requiring 
hypersurface--orthogonality, one additional scalar 
field is needed), but they did it only from the point of view how
to simplify the search for exact 4-dimensional solutions; 
they did not give the relation to dilaton gravity in 
2 dimensions.

In [25, 26], Geroch discussed the solutions of
Einstein's equation in the presence of one or two
 Killing vectors. Especially, in appendix A of [26] he gave a
 general construction to the reduced 2-dimensional 
formulation if
two commuting Killing vectors exist. 
     The main result of [25, 26] was the construction
of new solutions of the Kerr--Schild type 
$$\hat g_{ij} = g_{ij} + v_i v_j$$
from a given one $g_{ij}$. 
 This ``hidden symmetry'' of Einstein's equation 
has been applied in [27, 28, 29] for quantization issues. 
Both [27] and [28] use also a reduction of the 
4-dimensional Einstein equation with two
independent Killing vectors to a two-dimensional model, but
the details are quite different, so that no direct
comparison can be made. More detailed: [27] studies
the stationary axisymmetric case with the Ernst equation, and
[28] uses the complex Ashtekar variables.

Now let us return to the {4}-dimensional metrics of type eq. 
 $(1)$ possessing two commuting hypersurface-orthogonal
 space-like Killing vectors. These Killing vectors represent the
translations into $x$- and $y$-direction. This space-time is
called plane-symmetric if it possesses also
 the rotations in the
$x$-$y$-plane as isometries. Clearly, this is the case if and
only
if $\psi$ in eq. $(1)$ represents a constant.

According to $[19, 20]$, 
plane-symmetric solutions of the Einstein
equation possess a further
symmetry. Now, let us give a new proof of this 
statement using the results of {2}-dimensional gravity (the
analogous procedure for spherical instead of plane symmetry has
been carried out in ref. $[17, 18]$): For $n=2, D=4$, eq. $(8)$
reduces
to
\be
 L= (P- \frac{1}{2}\Box \phi - \frac{1}{2} \psi_{,\alpha}
\psi^{, \alpha}) e^{\phi} \sqrt{g} \ee
For $\psi=$ const., $L$ eq. (12) has the requested structure as
dilaton-metric theory without extra scalar field, and so we may
apply the
generalized Birkhoff theorem mentioned
 at the end of sct. $2$ giving
rise also to an additional isometry for the solutions
(1) if $\psi=0$. Of course, here it repeated only a known result,
but
this shall show how the method can be applied. The 
potential applications will induce the following: Let the
dilaton-metric $2$-dimensional theory with one additional scalar
field 
$\psi$ ($L$ eq. (12)) possess a certain class of solutions, then
they
give rise to analogous solutions of the $4$-dimensional Einstein
equation 
metric $(1)$, and all such solutions - 
including gravitational waves
 - will be reached by this procedure. The next step of
application
might be that any satisfactory quantization of 
$L$ eq. (12) may be directly transformed to a corresponding
quantization of $4$-dimensional gravity (cf. also [11] for
this).

The correspondences of this type discussed up to now
 had been essentially restricted to 
spherically 
symmetric gravity in $4$ dimensions, just excluding the 
gravitational waves from the beginning. But quantization of
gravity should include gravitational waves. We circumvented this
problem by changing from spherical symmetry to the 
symmetry of eq. $(1)$.

To elucidate the formalism at a
 concrete example, let us now consider a special class of
 pp-waves (plane-fronted waves with parallel rays) which can be
written in the form of eq. $(1)$ with vanishing $M$, i.e.,
 $d\sigma^2$ is flat, and the gradients of $\phi$ and $\psi$ are
parallel and lightlike. With these additional assumptions, eq.
$(1)$ may be written as
\be
ds^{2} = 2du dv - e^{\phi (u)}[e^{\psi(u)} dx^{2} + 
e^{-\psi(u)} dy^2] \ee 
It holds: Eq. $(13)$ represents a gravitational wave, i.e., the
Ricci tensor of $ds^2$ vanishes, if and only if 
\be \phi '{}' +
\frac{1}{2} \phi ' {}^{2} +
\frac{1}{2} \psi ' {}^{2}  = 0 \ee
where the dash denotes $\frac{d}{du}$. (And it is non-flat iff
 additionally $\psi '{}' + \phi ' \psi ' \ne 0$.)

On the other hand, if we insert $d \sigma^2 = 2du dv$ into the
field equation following from the $2$-dimensional Lagrangian $L$
eq.
$(12)$ and require that $\phi$ and $\psi$ depend on $u$ only,
then again just eq. $(14)$ remains to be solved. (By the way, it
follows already from the field equation that the gradients of
$\phi$ and $\psi$ are lightlike.) All the polynomial 
curvature invariants of metric $(13)$ can be expressed as 
polynomial invariants of the $2$-dimensional system $(12)$,
i.e., as
$\Box \phi$, $\Box \psi$, $\psi_{;jk} \phi^{;jk}$ etc.
The proof that all of them identically vanish can be directly
performed in the $2$-dimensional picture.


\section{Discussion}

One of the arguments why metric-dilaton gravity in two dimensions
should not be able to represent a 4--dimensional gravitational 
wave goes as follows:  
 The scalars in the 2--dimensional picture 
have spin zero, and so the spin 2--graviton
cannot be correctly incorporated by them. 

This argument can be outruled as follows: 
If one restricts the calculation from eq. (4)
 via eqs. (8), (9) till eq. (10) to $n=2$, then
one can see that the scalar $\psi$ of eq. (1) is a scalar in the 
2-dimensional point of view only.
 In the 4--dimensional picture one has to consider the
$\psi_1 - \psi _2 -$ plane
     (where $\psi =  2 \psi_1  = -  2\psi _2$) and
one has to consider the following part of the metric
$$
e^{\psi}dx^2 + e^{-\psi} dy^{2}
 = 
e^{2\psi_1}dx^2 + e^{2\psi_2} dy^{2}
$$
One can see: A rotation in physical space in the
$x-y-$plane by $90^0$ corresponds to 
$\psi_1 \longrightarrow - \psi_1$
and
$\psi_2 \longrightarrow - \psi_2$;  this 
represents a rotation by $180^0$ in the 
$\psi_1 - \psi _2 -$ plane. So the
spin is calculated as $180^0 / 90^0 = 2$, and the
correct tensor representation 
is maintained for the spin 2-gravitational wave.

Let us finish by mentioning 
some of the
results of refs. 
[3 - 12]: They discuss exact classical solutions
of dilaton-metric theories of gravitation $[6]$, of theories with
$L=P^{k}\sqrt{g}$ $[3]$, and of theories with torsion $[5,7]$.
In $2$ dimensions we have the peculiarity that the torsion tensor
$T_{\beta \gamma} ^{\alpha}$ is 
equivalent to the pseudovector field 
$$T ^{\alpha}=T_{\beta \gamma} ^{\alpha} \varepsilon ^{\beta 
\gamma}$$
where 
$\varepsilon ^{\beta \gamma}$ is the covariantly constant 
antisymmetric unit-pseudotensor. (Of course, the
 interpretations may differ.)

A lot deal of work has already be done to determine the global
behaviour, their horizons and singularities of the
 corresponding solutions, cf. [3,4,8,9,10,11,15]. In $2$
dimensions, the behaviour of the space-time is much easier to
understand
because the space-time is flat if and only if the curvature
scalar vanishes, whereas in $4$ dimensions (e.g. a
 gravitational wave of the form of eq. (1)) 
 a space-time 
 may be non-flat with
all polynomial curvature invariants being identically zero.  
By field redefinitions and/or conformal transformations many
 of the 
variants of $2$-dimensional gravity theories become 
equivalent (every paper [3-16] uses such transformations). For
theories $L=F(P)\sqrt{g}$ with non-linear $F$ and, 
according to this equivalence, also for dilaton-metric 
theories without extra scalar fields, i.e.
$$L=e^ \phi [P-\Box \phi +V (\phi)]\sqrt{g}$$
 a generalized Birkhoff theorem holds (deduced in $[16]$ for
$L=F(P)\sqrt{g}$ and in $[4, 30]$ for the other cases): every
solution possesses a symmetry.

The results presented here should motivate to direct future
research of 2--dimensional models to such Lagrangians which
have really a known 4-dimensional counterpart. 

\bigskip

\section*{Appendix on hidden symmetries}

Hidden symmetries of $2-$dimensional models are  discussed in
$[27-31]$ from several points of view. Geometrically, they have
three origins: The local conformal flatness of all $2$-spaces;
the fact, that locally, $P$ is a divergence; and, the fact that
the traceless part of $P_{ij}$ vanishes. 
The latter has already been
mentioned to  be the origin of the Birkhoff-type
theorem.
It should be mentioned that for the conformal equivalence of 
$L = f(P) \sqrt{\vert g \vert}$ and 
$\hat L = \phi \hat P - V(\phi)$ the necessary conformal
factor is a well-defined function of the curvature scalar $P$,
so this conformal equivalence of theories is {\it {\bf not}} a
consequence of the overall conformal flatness.

For any constants $a,b$ with $a \ne 0$ let
\be
 f_{a,b}(P) \ = \ a f(P) + b \,  P
\ee 
In this picture, the
``new-conformal extra symmetry''
 introduced in $[31]$ is simply a
variation of the constants $a$ and $b$. Here we will show that
even within the set of classical vacuum 
solutions\footnote{Of course, every element of the 2-parameter
family (15) of Lagrangians gives rise to the same
set of vacuum solutions}, this symmetry
is not so innocuous 
 as it seems from the first glance. To this end
we define $f^{(\epsilon)}(P) = P^{1+\epsilon}$ and ask for 
the limiting 
 theory as $\epsilon \to 0$.

If always $a=1,\ b=0$, then we get 
$$\lim_{\epsilon \to 0} f^{(\epsilon)}(P) = P$$
 If we put,
however, $a = \frac{1}{\epsilon}$, 
$b= - \frac{1}{\epsilon}$,
i.e., 
$f_{ab} ^{(\epsilon)}(P) = 
\frac{1}{\epsilon}(P^{1+ \epsilon} - P)
$ 
then 
$$
\lim_{\epsilon \to 0} f_{ab} ^{(\epsilon)}(P) = P \ln P
$$
 This means: If we factorize the set of Lagrangians with 
respect to this $a-b-$symmetry (15),
 then it is no more a Hausdorff space.

From a more geometric point of view this can be explained as
follows:
$P \sqrt{-g}$ in two dimensions gives no contribution to the
field
equation from two different reasons:
 First, $P \sqrt{-g}$ is locally a divergence. 
Second, $\int P \sqrt{-g}$  is a 
conformally invariant action, and
$2-$spaces are locally conformally flat. [On the contrary in four
dimensions, no polynomical curvature invariant exists which 
is simultaneously 
a local divergence and which gives rise to a
conformally invariant action.] 
And the coincidence of these two
properties produces the additional $P \cdot \ln P$
similar to ordinary linear differential equations where the
solution  $e^{\lambda x}$ is accompanied by
 $e^{\lambda x} \cdot x$ if the eigenvalue $\lambda$
 is a double one.

\bigskip

\noindent 
{\large {\bf Acknowledgement.}}

\medskip 

\noindent 
I thank H. Goenner, 
 V. Ivashchuk, V. Melnikov,
 M. Rainer and the referee for valuable comments and 
the Deut\-sche For\-schungs\-gemein\-schaft 
DFG for financial support. Further I thank the
Russian Gravitational Society in Moscow 
(where this paper has been finished) for
kind hospitality.

\bigskip

{\Large {\bf     References }}

\medskip

\noindent 
1.  J. Bicak and J. Griffiths, {\it Ann. Phys. NY}
 {\bf 252}, 180 (1996).

\noindent 
2. V. Gavrilov, V. Ivashchuk, U. Kasper and V. Melnikov, 
{\it Gen. Relat. Grav.} {\bf 29}, 599 (1997).

\noindent  
3.  S. Mignemi and H.-J. Schmidt, Preprint 
   Cagliari INFNCA-TH 9708 (1997),
 Classification of multidimensional inflationary models.
J. Math. Phys. in print. gr-qc/9709070.

\noindent 
4.   M. Cadoni, {\it Phys. Rev. D} {\bf 53}, 4413 (1996).

\noindent 
5.  M. Katanaev, W. Kummer and  H. Liebl, 
 {\it Nucl. Phys. B} {\bf 486}, 353 (1997).

\noindent 
6. S. Mignemi, {\it Ann. Phys. NY} {\bf 245}, 23 (1996).

\noindent 
7. S. Mignemi, {\it Mod. Phys. Lett. A} {\bf 11}, 1235 (1996).

\noindent 
8.  T. Kl\"osch and T. Strobl, {\it Class. Quant. Grav.}
             {\bf 13}, 2395 (1996).

\noindent 
9.  T. Kl\"osch and T. Strobl, {\it Class. Quant. Grav.}
             {\bf 14}, 1689 (1997).

\noindent 
10. C. Bernutat, thesis University Potsdam, $1996$.

\noindent 
11. J. Creighton and R. Mann, {\it Phys. Rev. D} 
   {\bf 54}, 7476 (1996).

\noindent 
12. M. Katanaev, {\it J. Math. Phys.} {\bf 38}, 946 (1997).

\noindent 
13. S. Solodukhin, {\it Phys. Rev.} {\bf D
51}, 
603 (1995).

\noindent
14. Y. Peleg, S. Bose, and L. Parker, {\it Phys. Rev.} {\bf D
55}, 
R4525 (1997).

\noindent 
15. M. Leite and V. Rivelles, {\it Phys. Lett. 
       B} {\bf 392}, 305 (1997).

\noindent 
16. H.-J. Schmidt, {\it J. Math. Phys.} {\bf 32}, 1562 (1991).

\noindent 
17.  S. Mignemi and H.-J. Schmidt, {\it Class. Quant. Grav.}
  {\bf 12}, 849 (1995).

\noindent 
18. H.-J. Schmidt,  $1997$: A new proof of
       Birkhoff's theorem, Grav. and Cosmology 3 (1997) 185; 
gr-qc/9709071.

\noindent 
19. H. Goenner, {\it Commun. Math. Phys.} {\bf 16}, 34 (1970).

\noindent 
20. V. Ruban, Abstr. Conf. GR $8$ Waterloo 1977, p. 303.

\noindent 
21. M. Rainer and A. Zhuk,
 {\it Phys. Rev.} {\bf D 54}, 6186 (1996).

\noindent 
22. F. Cooperstock and V. Faraoni, {\it Gen. Relat. Grav.} 
        {\bf 27}, 555 (1995).

\noindent 
23. B. Mann and A. Sikkema, {\it Gen. Relat. Grav.}  
       {\bf 27}, 563 (1995).

\noindent
24. D. Kramer, H. Stephani, M. MacCallum, E. Herlt: Exact
 solutions of Einstein's field equations, Verl. d. 
Wissenschaften Berlin 1980.

\noindent
25. R. Geroch, {\it J. Math. Phys.} {\bf 12}, 918 (1971).

\noindent
26. R. Geroch, {\it J. Math. Phys.} {\bf 13}, 394 (1972). 

\noindent
27. D. Korotkin, H. Nicolai, {\it Phys. Rev. Lett.}
 {\bf 74}, 1272 (1995).

\noindent
28. V. Husain, {\it Phys. Rev.} {\bf D 53}, 
4327 (1996).

\noindent
29. A. Ashtekar and M. Pierri, {\it J. Math. Phys.} {\bf 37}, 
6250 (1996).

\noindent
30. T. Banks, M. O'Loughlin, {\it Nucl. Phys. B} 
{\bf 362},
 649 (1991).

\noindent
31. J. Cruz, J. Navarro-Salas, M. Navarro, C. Talavera, 
{\it Phys. Lett. B} {\bf 402}, 270 (1997). 

\end{document}